\documentclass[a4paper,11pt]{article}
\usepackage{pos}
\usepackage{subfig,graphicx}

\title{Penguin decays of B mesons}

\author*[a]{Aritra Biswas}

\affiliation[a]{Theoretische Physik 1, Center for Particle Physics Siegen (CPPS), Universit\"at Siegen,\\
  Walter-Flex-Str. 3, 57068 Siegen, Germany}

\emailAdd{Aritra.Biswas@uni-siegen.de}

\abstract{We propose a set of optimized observables using penguin mediated $\bar{B}_d$ and $\bar{B}_s$ decays to $K^{(*)}\bar{K}^{(*)}$ and $K^*\phi$ final states that exhibhit a reduced sensitivity to power corrections than the corresponding branching ratios. Using these observables, branching ratios for the related vector-vector and pseudoscalar-pseudoscalar modes and branching ratios for modes involving other relevant pseudoscalar-vector final states ($K\phi$, $K^*\bar{K})$, we attempt to resolve the emergent combined pattern by assuming new physics contributions to the weak effective operators at the scale of the mass of the $b$ quark that are already present in the Standard Model. We find that the simplest scenario that can offer a simultaneous explanation is the two-operator scenario $Q_{4s,d}-Q_{6s,d}$. We briefly discuss how future experimental measurements may alter this picture and try to motivate such measurements for specific modes from our analysis.}

\FullConference{9th Symposium on Prospects in the Physics of Discrete Symmetries (DISCRETE2024)\\
 2–6 Dec 2024\\
Ljubljana, Slovenia\\}

\begin{document}
\begin{flushright}
	SI-HEP-2025-04
\end{flushright}

\maketitle

	\section{Introduction}
In a set of recent articles (refs.~\cite{Biswas:2023pyw,Biswas:2024bhn}), we looked at penguin mediated two-body non-leptonic decays of the $\bar{B}_{d,s}$ mesons into exclusive final states such that the underlying quark transitions are related via U-spin symmetry; using the framework of QCD factorization (QCDF) discussed in details in refs.~\cite{Beneke:2003zv,Bartsch:2008ps}. We showed that the U-spin relation can be exploited to construct ``optimised" observables called $L$, which are simply the ratio of branching ratios (or, in the case of purely vectorial final states, ratios of the longitudinally polarized branching ratios). These observables are optimised in the sense that they have a reduced sensitivity to the large power corrections that the corresponding branching ratios suffer from, and hence have a relative uncertainty that is less than the relevant branching ratios. Deviations at the level of $~2.5 \sigma$ and $1.5 \sigma$ were observed for these observables corresponding to $K^{(*)}\bar{K}^{(*)}$~\cite{Biswas:2023pyw} and $K^*\phi$~\cite{Biswas:2024bhn} final states respectively. We realized that the deviations for the vector-vector final states were being driven by the corresponding $b\to d$ branching ratios, whereas that for the pseudoscalar-pseudoscalar final states by the relevant $b\to s$ branching ratio. A more intricate pattern emerges if one considers related penguin dominated branching ratios to pseudoscalar vector final states corresponding to either a $b\to d$ or $b\to s$ transition \footnote{Optmised observables for these final-states cannot yet be constructed since they require the measurements of both $b\to s,d$ branching ratios for a particular final state. Such simultaneous measurements are currently unavailable in the literature for these modes.}. We looked into the possibility of a simultaneous explanation of the emergent pattern in terms of New Physics (NP) contributions to the operators already present in the weak-effective theory (WET) at the scale of the mass of the bottom quark in the Standard Model (SM).

This is a short proceeding highlighting the relevant observations and important results obtained and discussed in details in refs.~\cite{Biswas:2023pyw,Biswas:2024bhn}. The structure of the proceeding is as follows. In section~\ref{sec:th} we introduce and discuss the theoretical framework that was used for the construction of the optimised observables, as well as point out the important differences among the different modes considered. We provide the SM predictions calculated in QCDF, experimental measurements and the corresponding statistical deviations in section~\ref{sec:obs_val_SM_exp}. Section~\ref{sec:NP} is dedicated to a short discussion and presentation of the important results we find from the NP analysis. We finally conclude and discuss possible directions of relevant research in the future in section~\ref{sec:conclusions}.

\section{Theory}\label{sec:th}
The amplitude for a general two body non-leptonic decay of the pseudoscalar $\bar{B}_{p}$ ($p=s,d$) meson into two-body mesonic final states $M_1 M_2$ entailing the quark current $b\to q$ can be written as:
\begin{equation}
\bar{A}_f\equiv A(\bar{B}_p\to M_1M_2)
=\lambda_u^{(q)} T_{q} + \lambda_c^{(q)} P_{q}
=\lambda_u^{(q)}\, \Delta_{q} - \lambda_t^{(q)} P_{q} 
\label{dec}
\end{equation}
where $\lambda_U^{(q)}=V_{Ub} V_{Uq}^*$~\footnote{The weak phase in $\lambda_t^{(q)}$ is the angle $\beta_q$, defined as
	$\beta_q\equiv \arg \left(- \frac{V_{tb} V_{tq}^*}{V_{cb} V_{cq}^*} \right)= \arg \left(- \frac{\lambda_t^{(q)}}{\lambda_c^{(q)}} \right)\,,
	$} and $\Delta_q=T_q-P_q$. The corresponding CP-conjugate amplitude is:
\begin{equation}
A_{\bar{f}}=(\lambda_u^{(q)})^* T_q + (\lambda_c^{(q)})^* P_q    =(\lambda_u^{(q)})^* \Delta_q - (\lambda_t^{(q)})^* P_q\,.
\end{equation}
In the above, $T_q$ and $P_q$  do not correspond to ``Tree" and ``Penguin" contributions in general. These are simply matrix elements that can (in principle) be calculated in a particular dynamical framework. In QCD factorization (QCDF) they are expressed as an expansion in $\alpha_s$ upto $1/m_b$ suppressed terms that entail long distance effects and endpoint divergences\footnote{The corresponding formulas for the pseudoscalar-pseudoscalar and pseudoscalar-vector final states can be found in ref.~\cite{Beneke:2003zv} while those for the vector-vector final states can be found in ref.~\cite{Bartsch:2008ps}.}. Furthermore, the quantity $\Delta_q$ which is the difference between $T_q$ and $P_q$ is protected from infra red divergences. In light of the above observations, and as shown and discussed in refs.~\cite{Alguero:2020xca, Biswas:2023pyw, Biswas:2024bhn}, one can construct "optimized" observables (called $L$) involving penguin-dominated $B_{d,s}$ decays to the same final state. When $M_{1,2}$ are both vector mesons, a natural hierarchy occurs among the three possible polarizations that these final states can be in due to the left-handed structure of the $V-A$ interactions in the SM. The result is that QCDF can only reliably predict the longitudinally polarized modes for vector-vector final states. 

In this proceeding we consider a number of final states that the $B_{d,s}$ mesons can decay into. Detailed discussions regading these can be found in refs.~\cite{Biswas:2023pyw,Biswas:2024bhn}. We present the current experimental measurements for the branching ratios and the $L$'s in table~\ref{tab:obs_val}. A few comments regarding these modes are in order. There is only one penguin topology that contributes to each of the $B_{d,s}\to K^{(*)0}\bar{K}^{(*)0}$ branching ratios as is shown in fig.1 of ref.~\cite{Biswas:2023pyw}. However, two topologies can contribute to the $K^*\phi$ final states: one similar to the $K^{(*)}\bar{K}^{(*)}$ case and another that is not. This can clearly be seen from a comparison fig.1 of ref.~\cite{Biswas:2024bhn} with fig.1 of ref.~\cite{Biswas:2023pyw}. The $b\to s(d)$ transition for the $K^{(*)}\bar{K}^{(*)}$ involves the $B_{s(d)}$ initial state. However, this situation is reversed for the $K^*\phi$ final state. Furthermore. the meson that takes away the spectator quark is different for the two-different topologies contributing to the $\bar{B}_s\to K^*\phi$ transition. 

\begin{table}[h]
	\centering
	\begin{tabular}{|c|c|c|c|}\hline
		\textbf{Observable} & \textbf{SM (QCDF)} & \textbf{Experiment} &\textbf{Deviation} \\\hline
		$10^6\times\mathcal{B}(\bar{B}_d\to K^0\bar{K}^0)$&$1.09^{+0.29}_{-0.20}$~\cite{Biswas:2023pyw}&$1.21\pm 0.16$~\cite{ParticleDataGroup:2022pth,Belle:2012dmz,BaBar:2006enb}&$0.4\sigma$\\
		$10^7\times\mathcal{B}(\bar{B}_d\to K^{*0}\bar{K}^{*0})_L$&$2.27^{+0.99}_{-0.74}$~\cite{Biswas:2023pyw}&$6.04^{+1.81}_{-1.78}$~\cite{Alguero:2020xca,ParticleDataGroup:2022pth,LHCb:2019bnl,BaBar:2007wwj}&$1.8\sigma$\\
		$10^5\times\mathcal{B}(\bar{B}_s\to K^0\bar{K}^0)$&$2.80^{+0.89}_{-0.62}$~\cite{Biswas:2023pyw}&$1.76\pm 0.33$~\cite{ParticleDataGroup:2022pth,LHCb:2020wrt,Belle:2015gho}&$1.6\sigma$\\
		$10^6\times\mathcal{B}(\bar{B}_s\to K^{*0}\bar{K}^{*0})_L$&$4.36^{+2.23}_{-1.65}$~\cite{Biswas:2023pyw}&$2.62^{+0.85}_{-0.75}$~\cite{ParticleDataGroup:2022pth,LHCb:2019bnl}&$0.9\sigma$\\
		$10^6\times\mathcal{B}(\bar{B}_d\to \bar{K}^{*0}\phi)_L$&$4.53^{+2.16}_{-1.80}$~\cite{Biswas:2024bhn}&$4.96^{+0.31}_{-0.30}$~\cite{BaBar:2008lan,Belle:2013vat,LHCb:2014xzf,CLEO:2001ium}&$0.3\sigma$\\
		$10^7\times\mathcal{B}(\bar{B}_s\to K^{*0}\phi)_L$&$2.19^{+1.05}_{-0.94}$~\cite{Biswas:2024bhn}&$5.56^{+2.78}_{-2.27}$~\cite{LHCb:2013nlu}&$1.3\sigma$\\
		$\mathbf{L_{K^*\bar{K}^*}}$&$\mathbf{19.53^{+9.14}_{-6.64}}$~\cite{Alguero:2020xca,Biswas:2023pyw}&$\mathbf{4.43\pm 0.92}$~\cite{Alguero:2020xca,Biswas:2023pyw}&$\mathbf{2.6\sigma}$\\
		$\mathbf{L_{K\bar{K}}}$&$\mathbf{26.00^{+3.88}_{-3.59}}$~\cite{Biswas:2023pyw}&$\mathbf{14.58\pm3.37}$~\cite{Biswas:2023pyw}&$\mathbf{2.4\sigma}$\\
		$\mathbf{L_{K^*\phi}}$&$\mathbf{22.04^{+7.06}_{-4.88}}$~\cite{Biswas:2024bhn}&$\mathbf{8.80^{+6.07}_{-2.97}}$~\cite{Biswas:2024bhn}&$\mathbf{1.5\sigma}$\\
		$10^5\times(\mathcal{B}(\bar{B}_s\to K^{*0}\bar{K}^0)+c.c.)$&$0.83^{+0.50}_{-0.25}$~\cite{Biswas:2024bhn}&$1.98\pm0.28\pm0.50$~\cite{LHCb:2019vww}&$1.4\sigma$\\
		$10^6\times\mathcal{B}(\bar{B}_d\to\bar{K}^0\phi)$&$4.28^{+2.71}_{-1.50}$~\cite{Biswas:2024bhn}&$7.3\pm 0.7$~\cite{Belle:2003ike,ParticleDataGroup:2022pth,BaBar:2012iuj}&$1.3\sigma$\\
		$10^6\times\mathcal{B}(B^-\to K^-\phi)$&$4.67^{+2.98}_{-1.63}$~\cite{Biswas:2024bhn}&$8.8^{+0.7}_{-0.6}$~\cite{ParticleDataGroup:2022pth}&$1.5\sigma$\\
		$10^6\times\mathcal{B}(B^-\to K^{*-}\phi)$&$4.94^{+2.34}_{-1.91}$~\cite{Biswas:2024bhn}&$4.96^{+1.16}_{-1.08}$~\cite{BaBar:2007bpi,Belle:2005lvd,Belle:2003ike}&$0.05\sigma$\\\hline
	\end{tabular}
	\caption{SM (QCDF) predictions, experimental measurements and corresponding deviations (in $\sigma$'s) of all the different observables considered in this proceeding along with their respective references. The optimized $L$ observables are displayed in bold.}
	\label{tab:obs_val}
\end{table}	

\section{Standard Model predictions and Experimental measurements} \label{sec:obs_val_SM_exp}
We provide the SM predictions (calculated in QCDF), experimental measurements and the corresponding deviations for all the observables considered in this proceeding in table~\ref{tab:obs_val}.

Note that the deviation for the vector-vector final states is being driven by the corresponding $b\to d$ transitions, while that for the pseudoscalar-pseudoscalar final states by the relevant $b\to s$ transition.
We now aim to investigate whether there is a possible explanation for the picture that emerges from table~\ref{tab:obs_val} if one considers contributions from NP. To keep the analysis simple, we will only consider NP contributions to $b\to s,d$ operators in the WET at the $m_b$ scale that are already present in the SM.

\section{New Physics Analysis} \label{sec:NP}
\begin{figure}[ht]\centering
	\subfloat[]{\label{fig:c46d}\includegraphics[width=0.40\textwidth,height=0.30\textwidth]{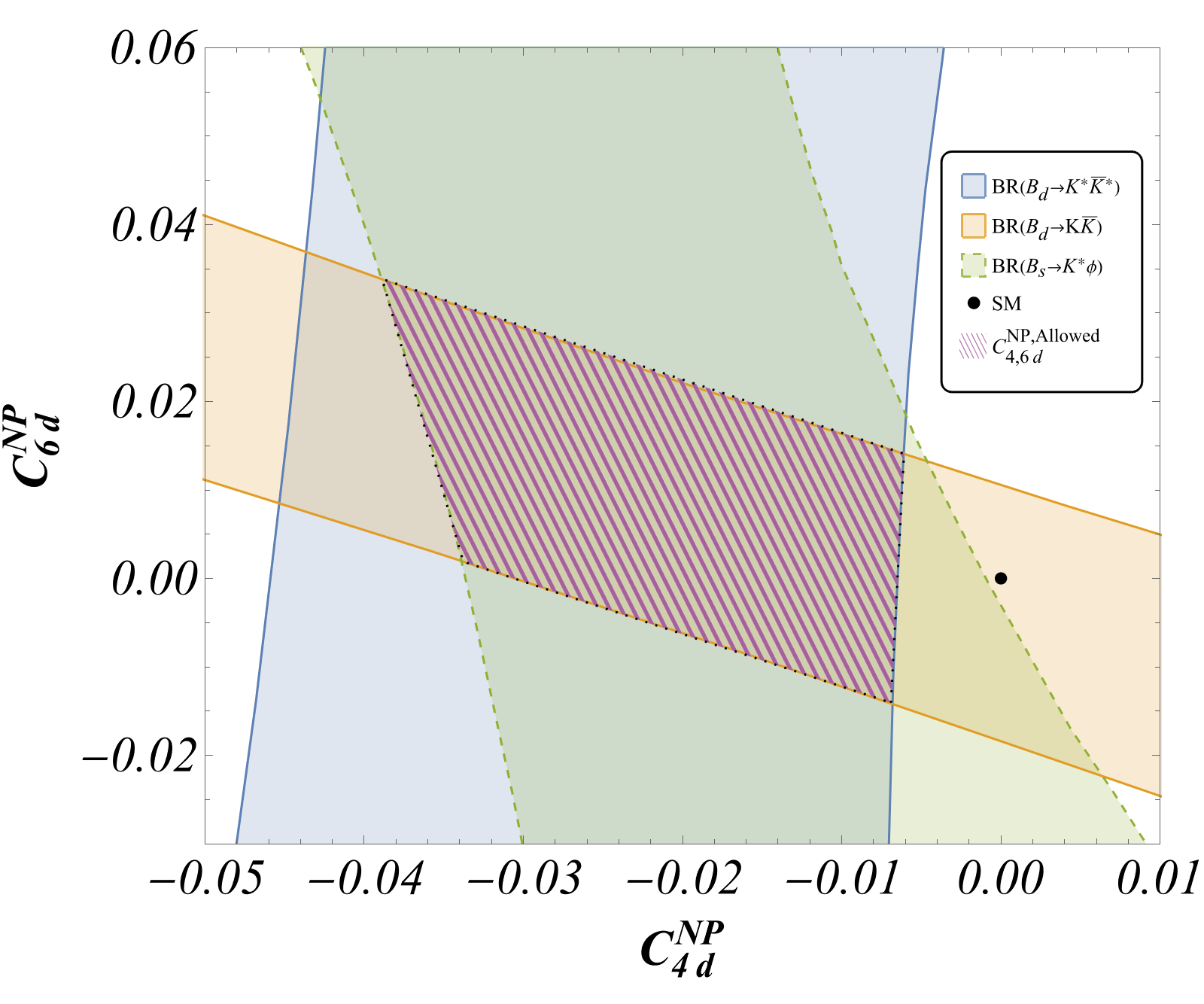}}~~~~~~~~~~~
	\subfloat[]{\label{fig:c46s}\includegraphics[width=0.40\textwidth,height=0.30\textwidth]{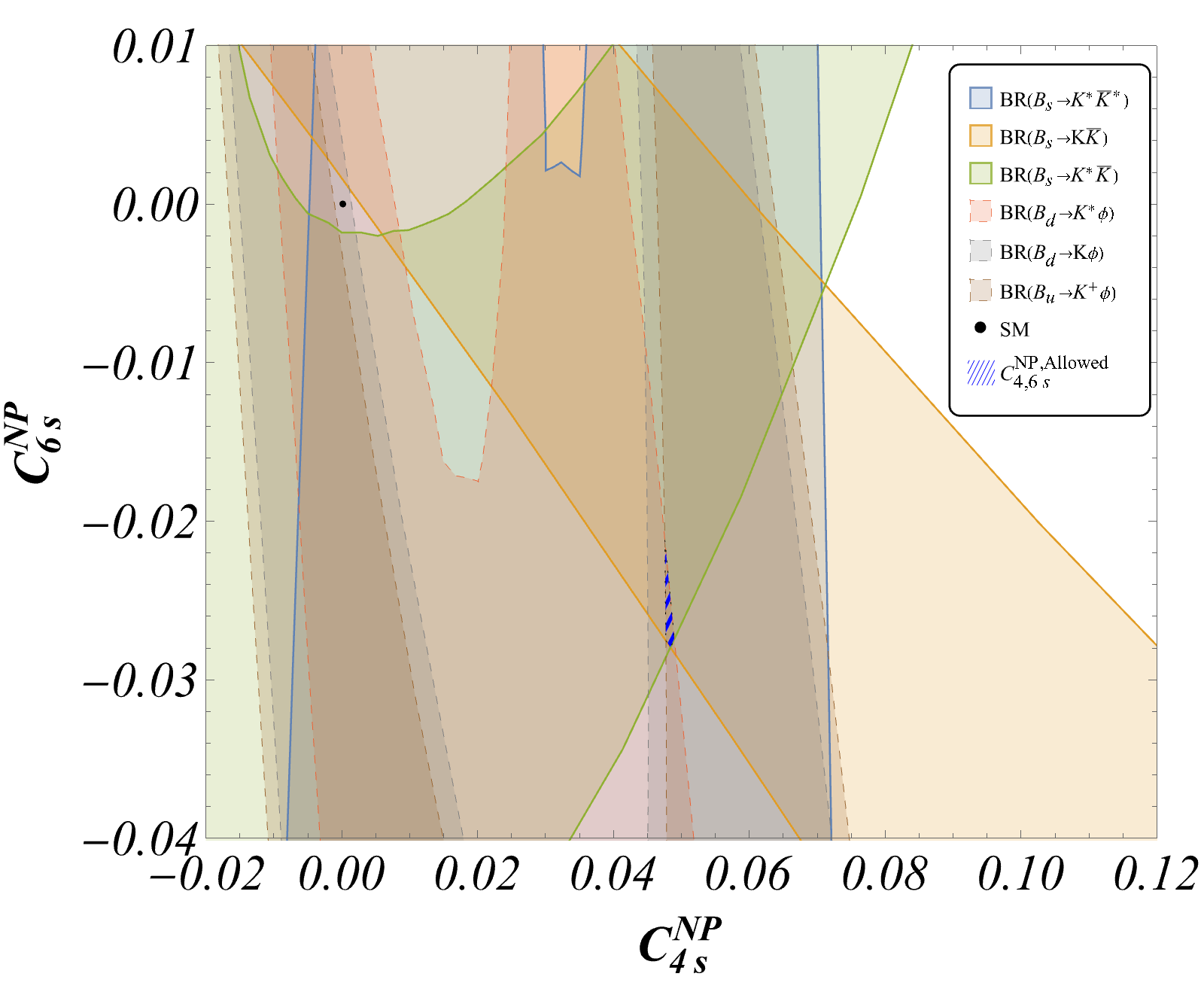}}\par 
	\subfloat[]{\label{fig:c46sz}\includegraphics[width=0.40\textwidth,height=0.30\textwidth]{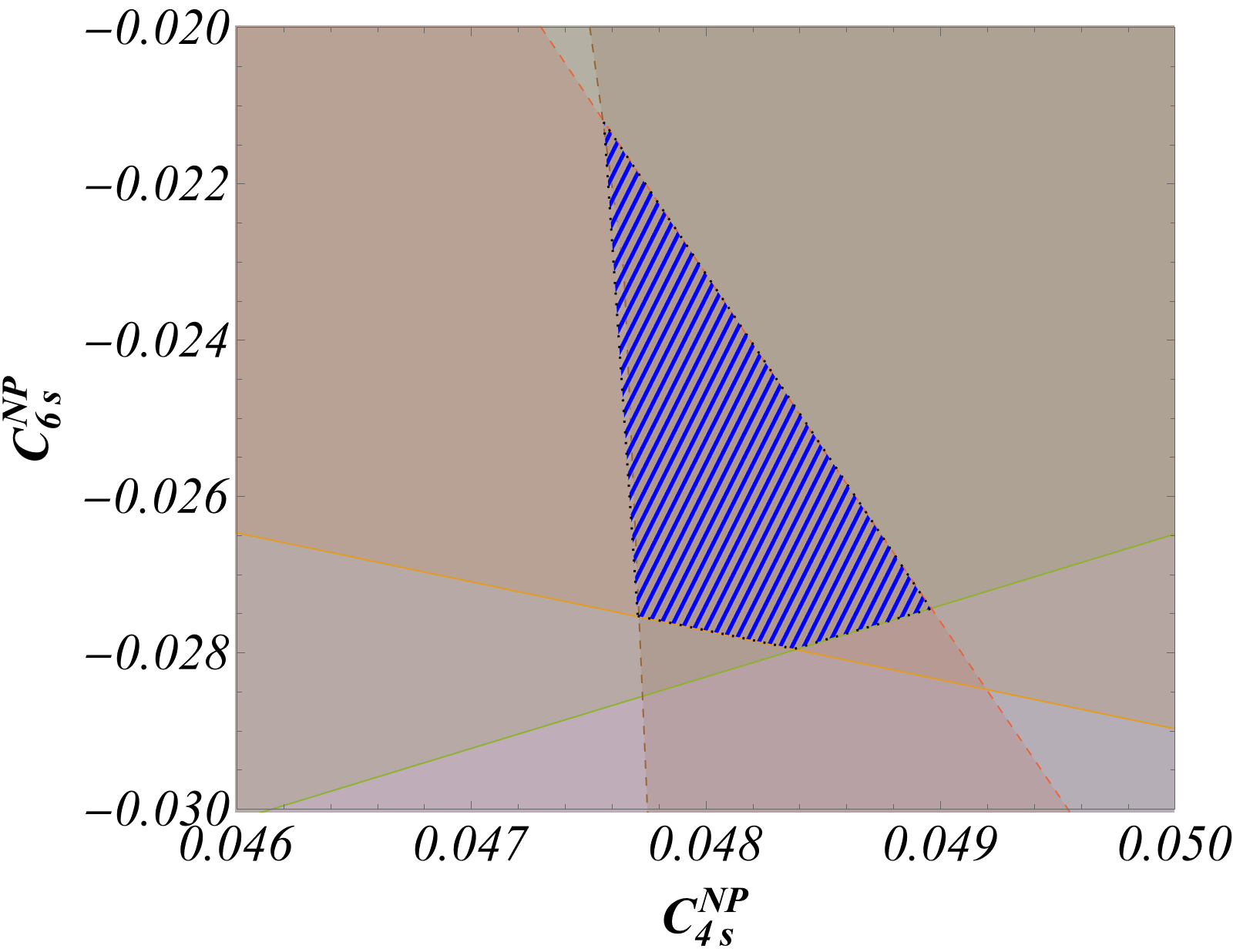}}		
	\caption{Parameter space for the two-operator $Q_{4s,d}-Q_{6s,d}$ scenario that can simultaneously explain all the ten branching ratioss provided in table~\ref{tab:obs_val}. Figs.~\ref{fig:c46d} and~\ref{fig:c46s} show the respective parameter spaces corresponding to the $b\to d$ and $b\to s$ branching ratios respectively, whereas fig.~\ref{fig:c46sz} is the zoomed-in version for the common region depicted in fig.~\ref{fig:c46s}. The $B^-\to K^{*-}\phi$ mode does not add any extra constraint to fig.~\ref{fig:c46s}. Hence we refrain from showing the corresponding band for aesthetic purposes.}
	\label{fig:c4c6br}
\end{figure}
The relevant hamiltonian for a general $b\to s,d$ transition at the scale $m_b$ is provided in appendix A of  ref.~\cite{Biswas:2024bhn}. In terms of the operator structure depicted there, the operators relevant for the simultaneous explanation of all the thirteen observables displayed in table~\ref{tab:obs_val} are $Q_{4s,d}$, $Q_{6s,d}$ and $Q_{8g,s,d}$. We begin by looking into ``one-operator scenarios", which essentially assume NP contributions to the corresponding Wilson Coefficients $C_i^{s,d} (i=4,6,8g)$ taken one-pair at a time. We find that such scenarios are incapable of providing a simultaneous explanation for all the observables. The operator $Q_{6d,s}$ does not work because they do not affect the pure vectorial final states in an influential way. The job of a simultaneous explanation of all the vectorial and pseudoscalar final states is achieved by the $Q_{4d,s}$ and $Q_{8gd,s}$ scenarios. However, this situation breaks down when the pseudoscalar-vector final states are included. 
\begin{figure}[h]\centering
	\subfloat[]{\label{fig:exp}\includegraphics[width=0.40\textwidth,height=0.30\textwidth]{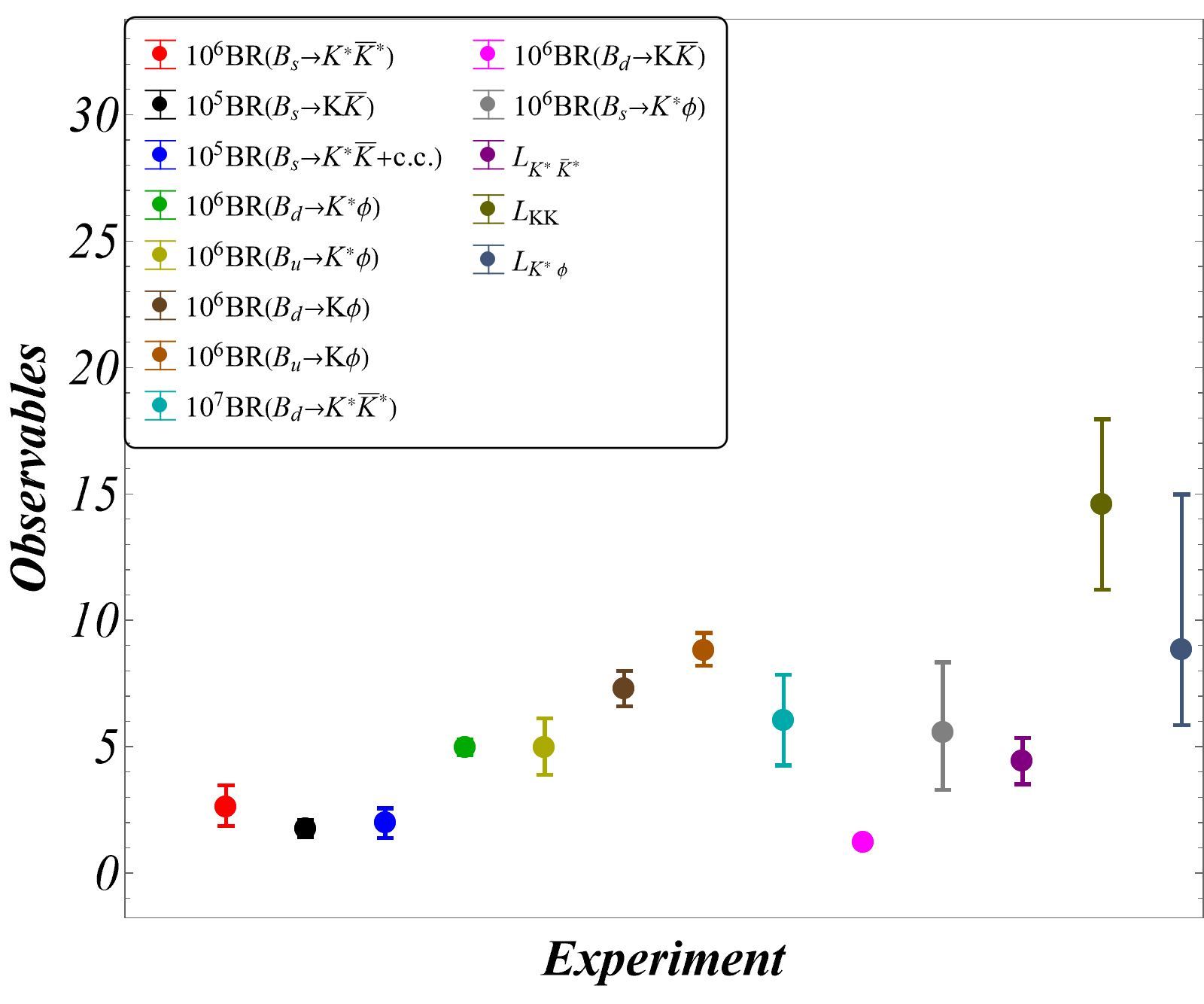}}~~~~~~~~~~~
	\subfloat[]{\label{fig:sm}\includegraphics[width=0.40\textwidth,height=0.30\textwidth]{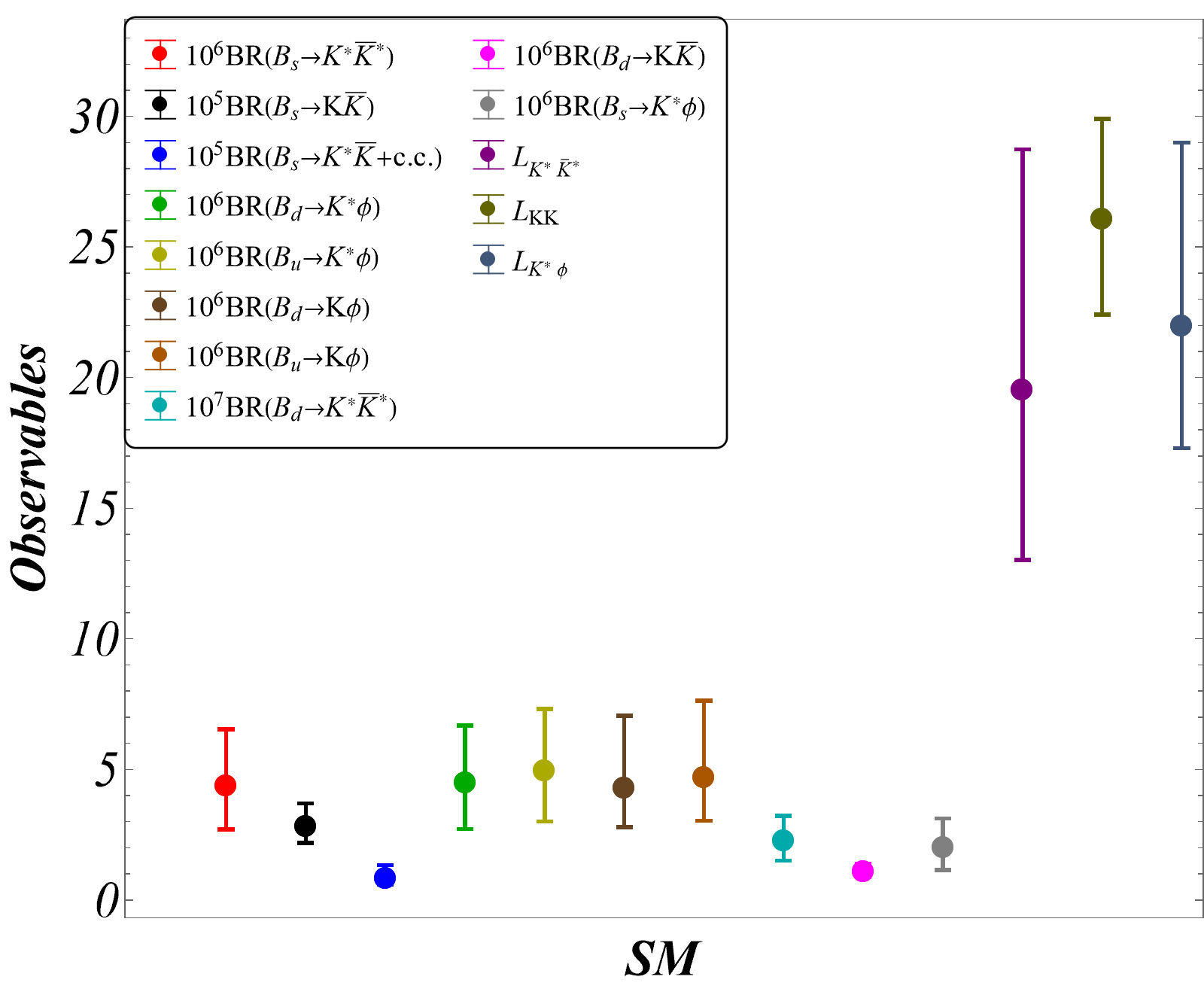}}\par 
	\subfloat[]{\label{fig:c46}\includegraphics[width=0.40\textwidth,height=0.30\textwidth]{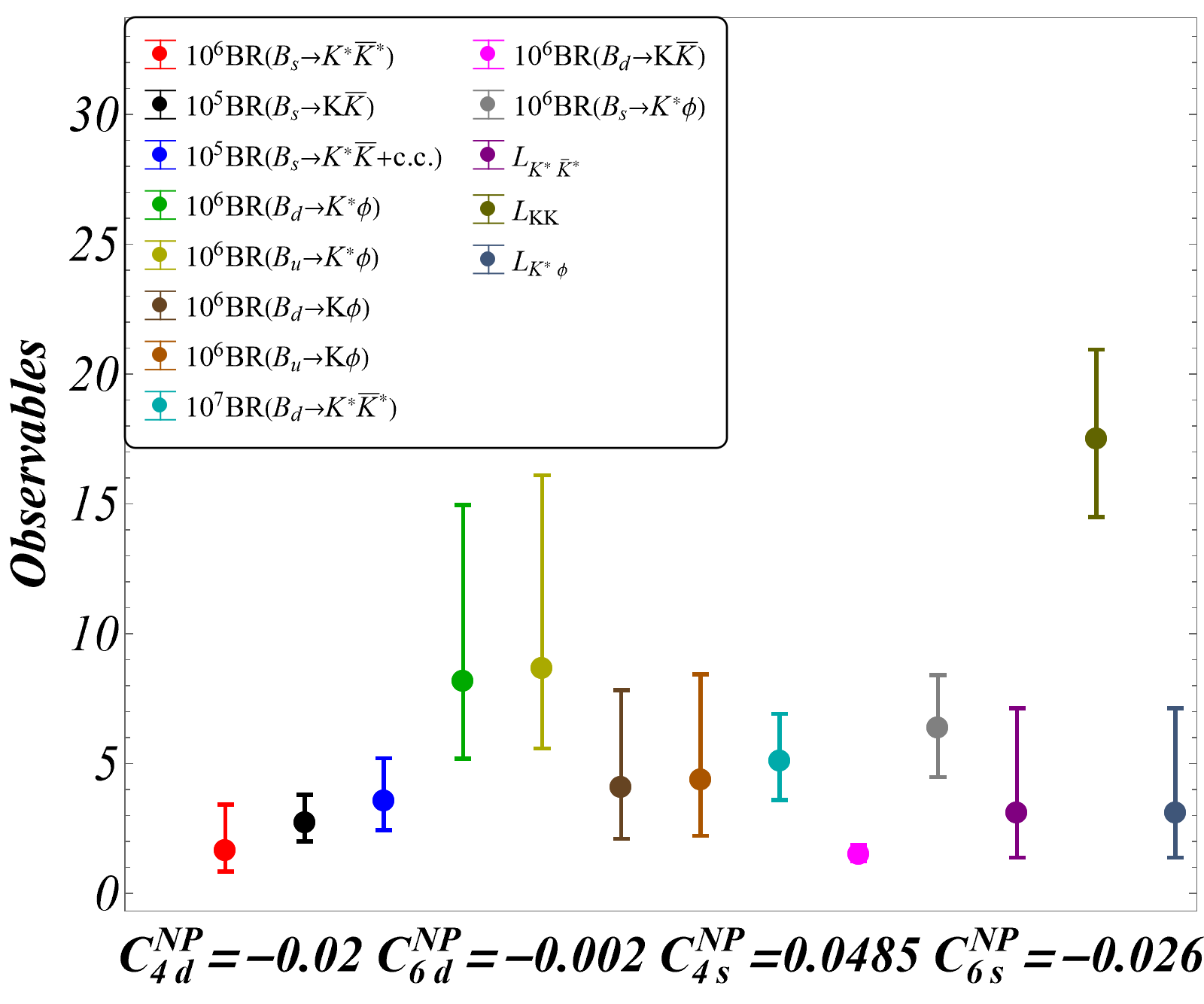}}		
	\caption{Experimental measurements and theoretical predictions of the thirteen observables (three $b\to d$ branching ratios, seven $b\to s$ branching ratios and three optimised observables $L$) in the SM and for benchmark values in the two-operator scenario $Q_{4d,s}-Q_{6d,s}$. Fig.~\ref{fig:c46} is fully consistent with Fig.~\ref{fig:exp} within $1\sigma$, contrary to fig.~\ref{fig:sm}. }
	\label{fig:npplt}
\end{figure}

The next logical step is to look at ``two-operator" scenarios $Q_{id,s}-Q_{jd,s}$, $(i,j=4,6,8g, i\neq j)$. To do this, we first scan the three $b\to d$ and seven $b \to s$ branching ratios separately. We find that only the $Q_{4d,s}-Q_{6d,s}$ scenario does the job. The resulting plots are shown in fig.~\ref{fig:c4c6br}, with the common regions of overlap (signifying the region in the parameter space that offers a simultaneous explanation for the corresponding branching ratios) marked with hatches. We then obtain benchmark points from both the common regions and test whether all or a subset of them also help in resolving the deviations observed in the three $L$'s. We find that they indeed do, and present a benchmark case for a particular instance of the possible $(C_{4d},C_{6d},C_{4s},C_{6s})$ combinations that provides a simultaneous explanation for all the thirteen observables in fig~\ref{fig:npplt}.

\section{Conclusion and Future directions} \label{sec:conclusions}
In this proceeding, we have discussed ``optimized observables" in non-leptonic $B_{d,s}$ decays, that can be constructed as ratios of branching ratios of $b\to s$ to $b\to d$ penguin dominated two-body final states. These observables display reduced uncertainties w.r.t the branching ratios used to construct them because they are less sensitive to power corrections owing to a protective U-spin symmetry between the numerator and the denominator. In the SM, the dominant sources for their uncertainties are the corresponding form factors. The patterns and deviations one observes between the SM predictions and experimental measurements, along with the corresponding (measured) branching ratios taken together may have a possible explanation in the form of NP contributions to the operators $Q_{4s,d}$ and $Q_{6s,d}$ considered together.

However, this scenario might be altered in the future. In particular the measurements for the charged and neutral $B\to K\phi$ branching ratios are more than a decade old. Their are no measurements from LHCb for these modes either. The current global averages for these modes as reported in ref.~\cite{ParticleDataGroup:2022pth} are $1.5\sigma$ apart, which is surprising since these modes are related by isospin. A careful measurement of these modes is hence warranted at this point of time. In particular, these two modes being consistent at $1\sigma$ will enable another two-operator scenario ($Q_{6d,s}-Q_{8gd,s}$) to provide simultaneous explanations for all the thirteen observables. On the other hand, a change in the measured value of the branching ratio for the $\bar{B}_d\to\bar{K}^0\phi$ might result in simpler one-operator scenarios being possible contenders for a simultaneous explanation as well. 

Correlated estimation(s) of the involved form factors will enable us to predict more robust uncertainties in the future. On the other hand, correlated experimental measurements will also help in providing a more comprehensive picture. Finally, we would like to stress that these are first exploratory works in this direction. We plan to undertake rigorous statistical analysis involving asymmetric distributions so that we can objectively infer the common regions overlap in terms of statistical quantities in the future.

\bibliographystyle{JHEP}

\bibliography{main.bib}



\end{document}